# Life cycle costing analysis of deep energy retrofits of a mid-rise building to understand the impact of energy conservation measures

Haonan Zhang[1]

*School of Engineering, Faculty of Applied Science, The University of British Coulmbia, Kelowna, BC, Canada, V1V1V7*

**Abstract**

Building energy retrofits have been identified as key to realizing climate mitigation goals in Canada. This study aims to provide a roadmap for existing mid-rise building retrofits in order to understand the required capital investment, energy savings, energy cost savings, and carbon footprint for mid-rise residential buildings in Canada.

This study employed EnergyPlus to examine the energy performance of 11 energy retrofit measures for a typical multi-unit residential building (MURB) in Metro Vancouver, British Columbia, Canada. The author employed the energy simulation software (EnergyPlus) to evaluate the pre-and post-retrofit operational energy performance of the selected MURB. Two base building models powered by natural gas (NG-building) and electricity (E-building) were created by SketchUP. The energy simulation results were combined with cost and emission impact data to evaluate the economic and environmental performance of the selected energy retrofit measures.

The results indicated that the NG-building can produce significant GHG emission reductions (from 27.64 $tCO_2e$ to 3.77 $tCO_2e$) by implementing these energy retrofit measures. In terms of energy savings, solar PV, ASHP, water heater HP, and HRV enhancement have great energy saving potential compared to other energy retrofit measures. In addition, temperature setback, lighting, and airtightness enhancement present the best economic performance from a life cycle perspective. However, windows, ASHP, and solar PV, are not economical choices because of higher life cycle costs. While ASHP can increase life cycle costs for the NG-building, with the financial incentives provided by the governments, ASHP could be the best choice to reduce GHG emissions when stakeholders make decisions on implementing energy retrofits.

# 1 Introduction

Extensive use of fossil fuels and associated greenhouse gas (GHG) emissions have been identified as catalysts for climate change and associated environmental impacts [1]. In response to the increasing concern about climate change impacts due to anthropogenic activities, the government of Canada has established an ambitious emission reduction target, promising to reduce carbon emissions by 80% by 2050 compared to the 2005 level [2]. It is suggested that stationary combustion, transportation, and fugitive sources constitute 82% of the GHG emissions of the country [3]. Thus, Canada aims to reduce energy use in multiple sectors in order to reduce associated GHG emissions [2].

In recent years, the building sector has garnered greater attention for its need to reduce GHG emissions in Canada. According to national GHG inventory, Canadian buildings are responsible for 12% of the total national GHG emissions. Moreover, the residential building sector accounts for 11% of national energy use in 2017 [2]. Recognizing the importance of reducing energy use and emissions associated with the building sector, governments have introduced policies, standards, and design guidelines to improve building energy performance. For example, the British Columbia Energy STEP Code (BCESC) has been launched to put British Columbia (BC) on a path to meet the provincial target to make all new buildings "net-zero energy ready" by 2032 [4]. However, policy initiatives that promote energy efficiency of aging buildings are lacking. Poor energy performance of aging buildings is responsible for a significant portion of the increasing GHG emissions associated with the building sector [9].

In Canada, over 50% of residential buildings are more than 30 years old, and over 20% are older than 50 years or more [5]. Old buildings mostly use non-renewable energy sources. Moreover, the systems and materials have often deteriorated with the age of the old buildings [8], [9]. Therefore, these old buildings consume more resources leading to higher environmental impacts and emission footprints [5]. As the replacement rate of old buildings by new construction is only 1.0–3.0% per annum [8], the environmental impacts associated with the old building stock need to be reduced through retrofitting in order to achieve the emission reduction targets of Canada [9,10]. Previous studies have shown that building energy retrofits have significant energy and GHG emissions saving potential [11]. Furthermore, retrofits can also deliver other benefits, such as cost savings and enhanced thermal comfort [12,13]. Therefore, devising energy performance enhancement and retrofit strategies for Canadian buildings is a timely discussion.

While municipalities have developed many plans to achieve resiliency and emissions reduction targets, retrofitting existing buildings for net zero or near net zero emissions remains one of the most challenging parts of reaching the climate action targets due to the diverse portfolio, early adopter disadvantage, and economic barriers. Multi-unit residential buildings (MURBs) represent a large and growing share of the building stock in cities across Canada.

In British Columbia, the City of Richmond intends to undertake a deep energy building retrofit of a mid-rise building to identify and verify the potential impacts of a series of energy conservation measures (ECMs) on energy consumption and GHG emission savings. This study aims to provide a roadmap for existing mid-rise building retrofits in order to understand the required capital investment, approximate energy savings, corresponding energy cost savings, and carbon footprint for a mid-rise residential building deep energy retrofit.

## 2 Energy retrofit interventions

In general, building retrofit interventions can be classified under three categories, including demand side solutions, supply side solutions, and transformation of energy consumption patterns (i.e. human factors) [12]. Demand side solutions include strategies to reduce building heating and cooling load and other end-uses with energy retrofits and upgrades. Upgrades in building envelop insulation, air tightness, window insulation, heating ventilation and air-conditioning (HVAC) systems, hot-water unit, and appliances are some common focus areas in demand-side management [15,16]. Supply side solutions consist of renewable energy solutions such as solar photovoltaics (PV) and wind energy, which are recognized as alternative energy systems to generate electricity for buildings [17]. Supply side solutions have received much scrutiny in recent years with the increasing pressures to reduce the environmental impacts associated with energy use [15,16]. Transformation of energy consumption patterns is applying advanced control techniques or providing householders with building operation strategies to facilitate energy efficiency through behavior change [18]. This section discusses possible retrofit options, including the improvement of building envelope components, HVAC systems, and occupant behaviors and lighting systems, and renewable energy systems.

### 2.1 Upgrades in building envelope

In existing buildings, heat losses or gains through building envelopes affect the energy use and the indoor condition, and produce a significant amount of energy depletion. Therefore, retrofitting the external walls and roofs has a considerable impact on reducing building energy consumptions. This kind of retrofit measures should improve the thermal performance of buildings [19]. Common upgrades in building envelope components include wall insulation, roof insulation, and windows [20].

### 2.2 Upgrades in HVAC systems

Previous research shows that the most substantial energy saving potential can be achieved by improving the building HVAC systems [19]. Existing old buildings usually have less-efficient HVAC systems, especially those systems powered natural gas. In the recent years, the development of heat pump have presented great energy saving potential for existing old buildings. In addition, HVAC control has an important impact on energy management [21]. HVAC control aims to optimize operation systems and avoid excessive cycling of system components and the conflicts between them. Adjustments in HVAC control strategies are essential to reduce building energy consumptions [12].

### 2.3 Occupant behaviours and lighting systems

Multi-unit residential building (MURB) occupants without electrical sub-metering tend to use more electricity than those who are sub-metered [12]. The cost of electricity is often hidden in fixed costs, while the actual electricity cost per suite varies according to fluctuations in energy use. Individual sub-metering ensures occupants are aware of their energy consumption. In order to manage energy costs, occupants have the option to reduce their in-suite appliance use and lighting. Light-Emitting Diode (LED) lighting have presented greater energy saving potential compared to conventional lights [10].

### 2.4 Renewable energy systems

In addition to demand side retrofit measures, supply-side retrofit measures, such as building integrated solar PV systems, are becoming popular in Canada. Solar PV can generate and store electricity during daytime and this electricity is used when demanded. Majority of the installed solar PV panels are equipped with an inverter to convert DC to AC and allow the household to utilize the produced electricity [22]. In addition, solar PV can be connected to the electricity grid. If the supplied electricity to the grid is more than the used electricity, the credits will carry over to the next utility bill [23].

## 3 Building energy modeling

This research selected a typical mid-rise MURB located in the City of Richmond, British Columbia, Canada as a case study. The author employed the energy simulation software (EnergyPlus) to evaluate the pre-and post-retrofit operational energy performance of the selected MURB. EnergyPlus (developed by Lawrence Berkeley National Laboratory) is the most popular energy simulation and design tool for buildings across the world. It employs long-term monthly weather data in a bin-based method to analyze energy performance for a given building.

### 3.1 Base building definition

The selected building is a three-floor muti-unit residential building. The building floor plan is shown in Figure 1.

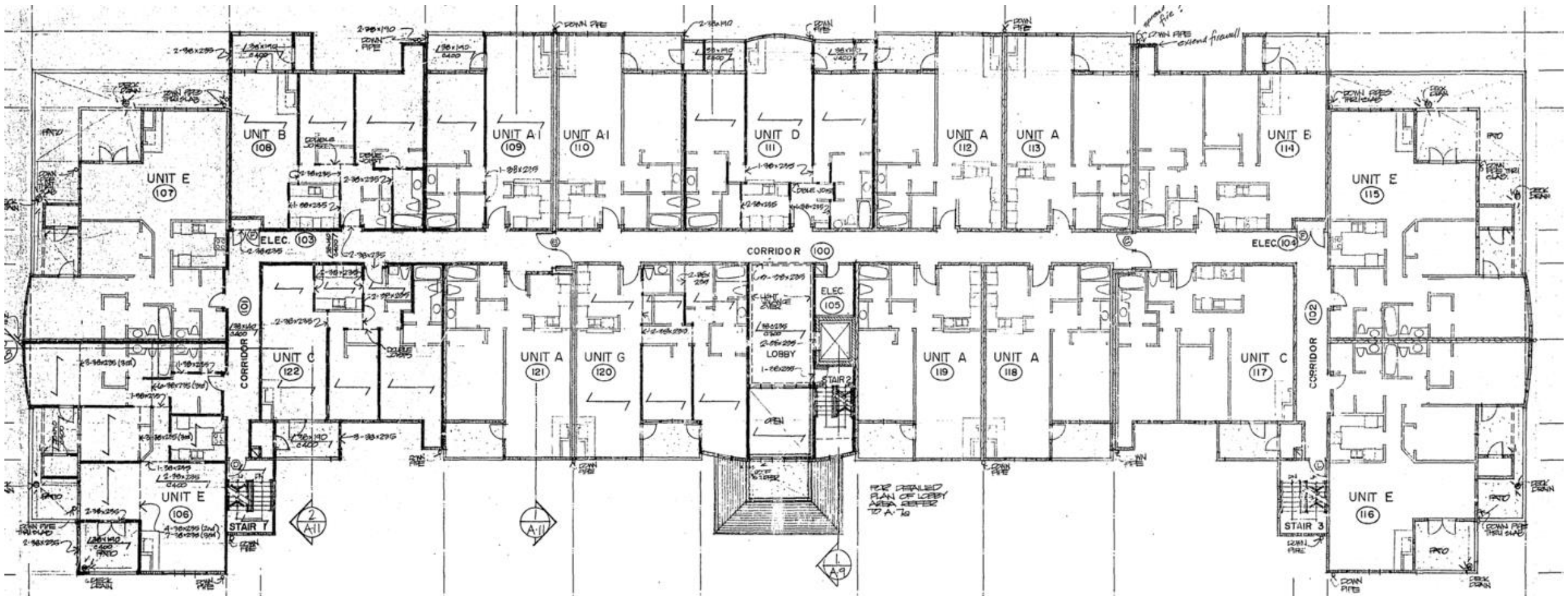

Figure 1. The floor plan of the case study building

### 3.2 Considered energy retrofit measures

#### 3.2.1 Envelope

**Add roof insulation:** R16. These base cases along with a series of roof insulation retrofits ranging from 2-inches (R-8) to a maximum of 4-inches (R-16)

**Add wall insulation:** R16. These base cases along with a series of roof insulation retrofits ranging from 2-inches (R-8) to a maximum of 4-inches (R-16)

**Windows improvement:** Triple pane, U value: 1.044 w/m2 K, SHGC: 0.615

**Air tightness improvement:** 0.001024 m3/s-m2

### 3.2.2 HVAC system:

**Space heating:** Air Source Heat Pump (ASHO): coefficient of performance: COP~2.75

**Water heating:** Water Heater Heat Pump (Water HP): COP~3.0

**Heating recovery ventilator (HRV):** sensible eff. 0.65

### 3.2.3 Renewable energy:

**Solar PV:** Solar PV can be installed on roof top to generate and store electricity during daytime and this electricity is used when demanded.

### 3.2.4 Occupant behaviors and others:

**Heating temperature setpoint setback:** This measure is a low/no cost measure that leads to saving a great deal of energy at hours that space temperature does not necessarily require to be at design temperature. While the set point temperature on occupied hours remains unchanged at 22, with this measure the setback temperatures for heating are adjusted to 18 °C.

**Lighting:** Conventional lighting are replaced with LED lights.

**Appliances:** Appliances are replaced with electric appliances (15% higher energy efficiency)

## 3.3 Energy modeling process

Typical MURBs in BC can be categorized into two clusters according to their energy sources: electricity-powered buildings and natural gas-powered buildings. To represent the majority of existing MURBs in BC and develop a comprehensive retrofit plan, the author developed two building energy models according to the energy sources.

The authors employed Google SketchUP with OpenStudio as an add-in to create the physical base building model, as shown in Figure 2. Then, the building model was imported to EnergyPlus to evaluate the energy performance of different retrofit measures.

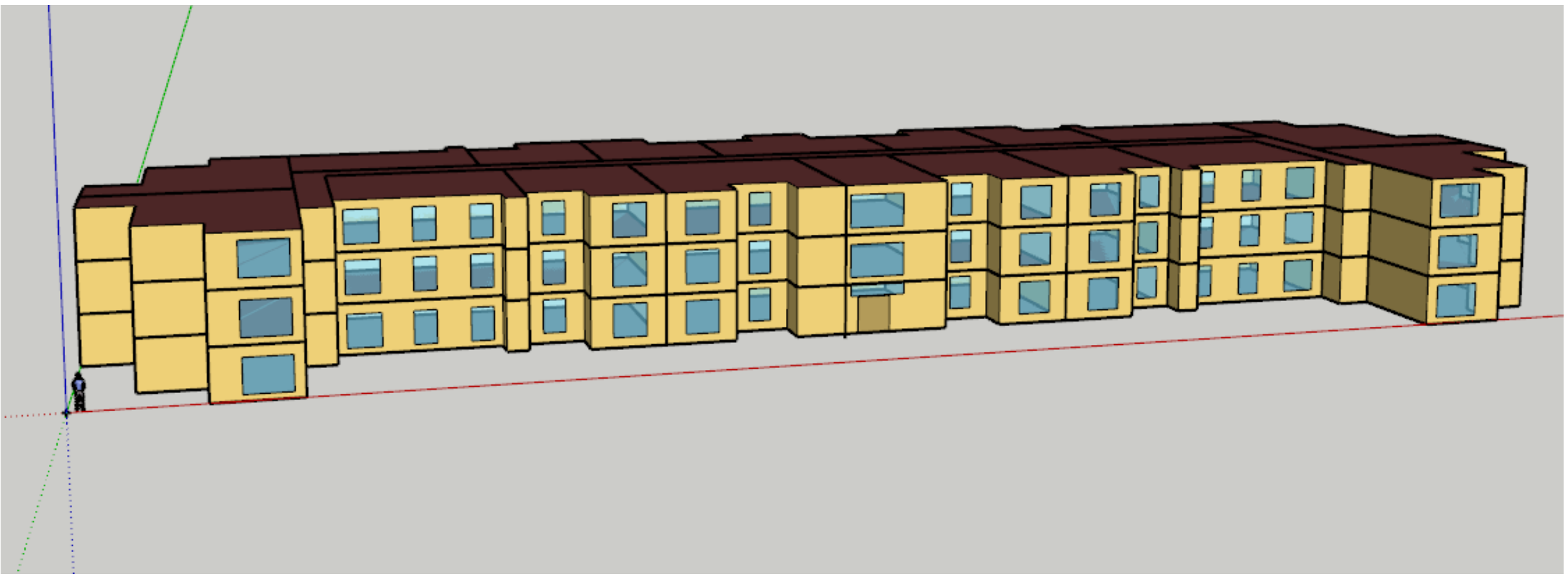

Figure 2. Physical base building model in SketchUP

After importing the physical building model to EnergyPlus, the author must input more information associated with building envelope materials, HVAC systems, lighting, electric equipment. The building characteristics are shown in Table 1. The developed building energy models are shown in Figure 3 and Figure 4. In order to evaluate the effectiveness of different retrofit measures, comparing the post-retrofit performance against the original performance of a given building is necessary. Thus, after creating the base building models, the author must input the selected energy retrofit measures in the base building energy models and evaluate the post-retrofit building energy performance. The pre- and post-retrofit energy performance is discussed in Section 5.

Table 1. Base building model

| Building characteristics | | |
|---|---|---|
| Specifications | Data | |
| HVAC System | Fan coil unit, Water cooled chiller, Boilers and chillers; Heating system seasonal NG CoP: 0.8, E CoP: 1.0, | |
| Domestic hot water | Electricity, Dedicated hot water boiler, Delivery temperature: 60°C, CoP: 0.85 | |
| Thermostat | Heating: 22°C, Cooling: 26°C | |
| Air infiltration | 0.001314 m3 /s-m2 | |
| Exterior Wall | Stucco, Wire mesh on building paper, 9.5 Plywood, 38*89 Studs, R-14 batt insulation, 12 Drywall | |
| Floor RSI | RSI=0.66 Carpet, plywood, joist, drywall | |
| Roof RSI | Ceiling under Attic | Roofing, 38*235 Joists, R-28 batt insulation, 16 Drywall |
| Windows and WWR% | U - value | 3.57 W/m2 K |
| | SHGC | Double glazed aluminum frame - 0.760 |
| Lighting density | 8.5250 W/m2 | |
| Plug load | 8.0729 W/m2 | |

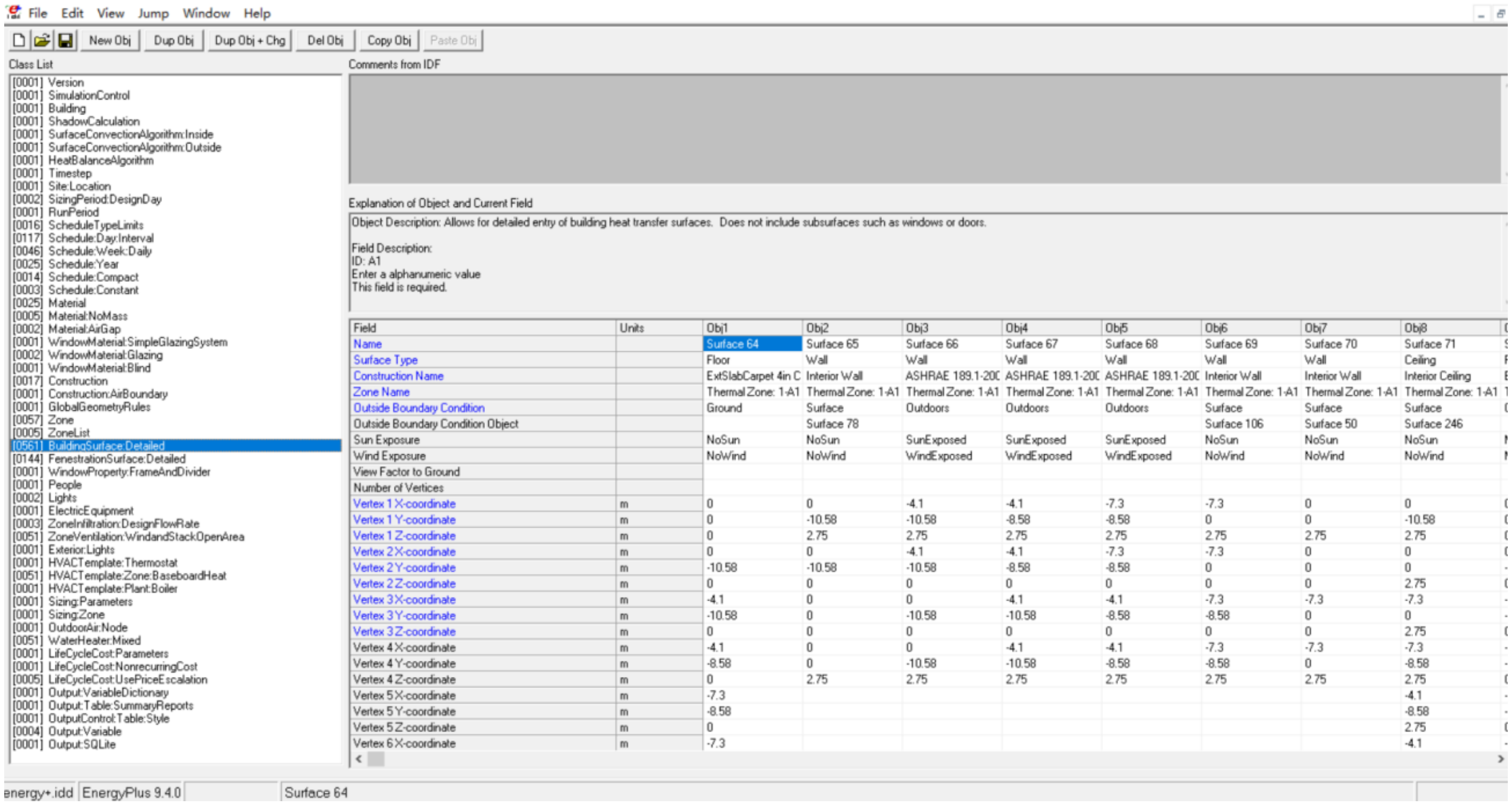

Figure 3. Electricity-heated building energy model

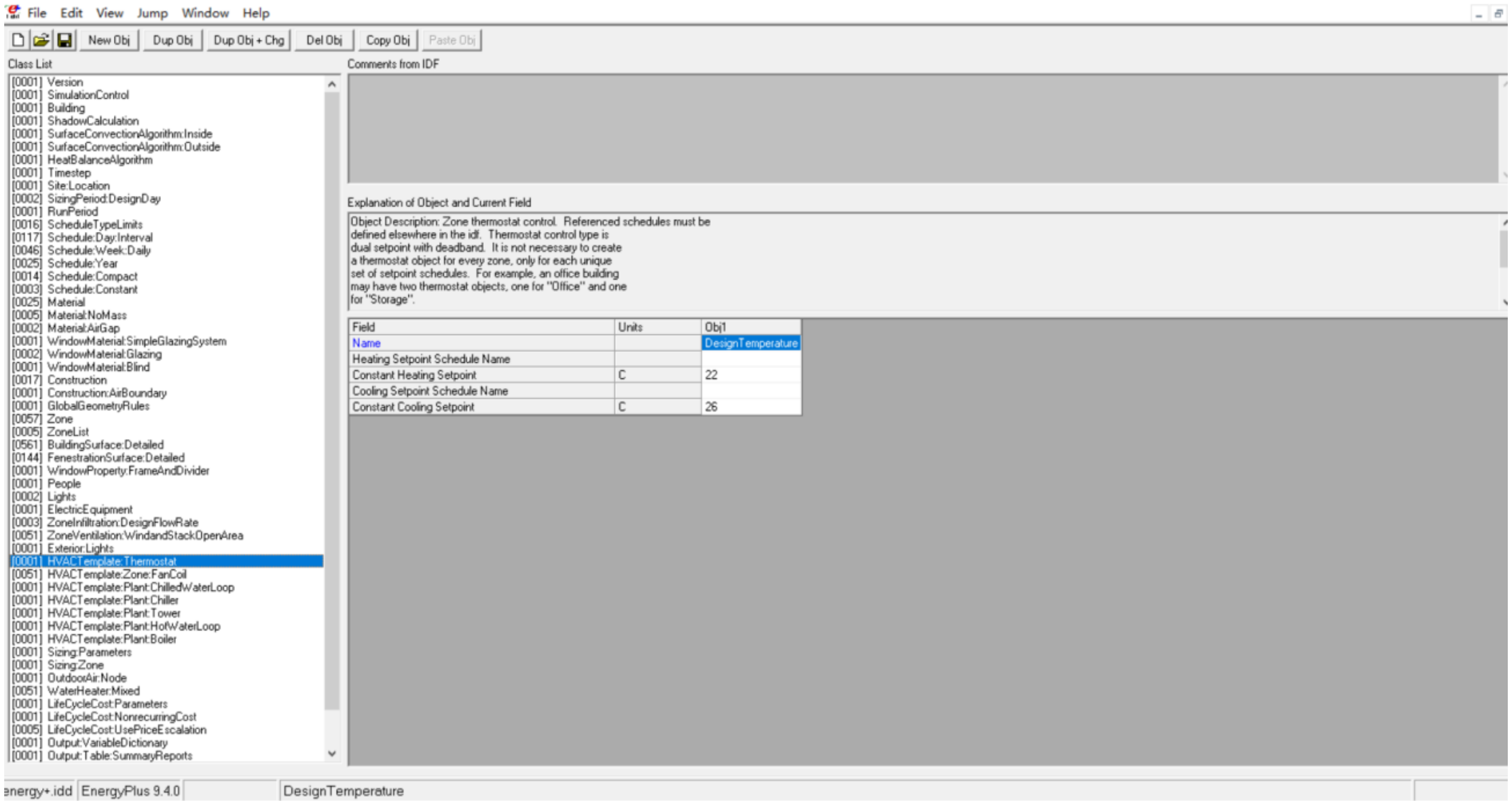

Figure 4. Natural gas-heated building energy model

# 4 Life cycle costing

The upfront cost of a retrofit project is an essential consideration for building owners. Homeowners are inclined to choose cheaper equipment or material to reduce the upfront cost, according to a retrofit survey conducted in Canada [24]. However, purchasing equipment or material with low market prices without considering the operational performance might raise the LCC. The LCC accounts for all cost elements associated with a retrofitting project. Depending on the conditions, a retrofit package with a higher upfront cost may produce better LCC performance due to higher cost savings [25].

Life cycle cost analysis (LCCA) is generally regarded as a pecuniary evaluation method for an existing asset or a potential investment. LCCA accounts for immediate and long-term expenses. LCC is the "cost of an asset or its parts throughout its lifecycle while meeting the performance requirements". In the building and construction sector, ISO 15686–5 was issued for the financial evaluation of "Buildings and con- structed assets". In this study, the considered LCC includes the upfront cost, the operational cost, and the disposal cost.

## 4.1 Upfront cost

Upfront cost is a combination of cost of equipment and installation. In this study, RSMeans Building Construction Costs database and literature were referred to identify the capital costs of the identified retrofits. For a given retrofit scenario, upfront costs (UC) associated with envelop and energy system upgrades can be calculated by the following equation.

$$UC = cc_{envelope,i} \times A_{envelope,i} + CC_{system,j}$$

Where,

- $cc_{envelope,i}$ = The unit capital cost of the $ith$ building envelope material
- $A_{envelope,i}$ = The area of the $ith$ building envelope component
- $CC_{system,j}$ = The capital cost of the $jth$ energy systems

## 4.2 Carbon tax

The energy cost rates of this building were ¢15 and ¢3.2 per kWh of electricity and natural gas respectively. An additional cost that needs to be considered atop the energy costs is Carbon tax costs. While the current carbon tax cost in Canada is $50 per ton of CO2e, based on the government's announcement, it is expected to increase by $15 per year, reaching $170 per ton in 2030. Although there could be a stop in carbon tax increase in 2030 at $170/ton, it is also reasonable to assume that the price of carbon will continue to increase. To date there is no official announcement for the carbon tax changes after 2030. However, some news are spread around increases to $300 by 2050. In this study, linear increase of carbon tax after 2030 until it reaches $300 by 2050 was assumed.

$$\Delta CTS = \left[\left(EE_{base} - EE_{retrofitted}\right) * EF + \left(NE_{base} - NE_{retrofitted}\right) * NF\right] * CT$$

Where,

- $\Delta CTS$ = The annual carbon tax cost savings (CAD)
- $EE_{base}$ = The annual electricity consumption of the base building model (GJ)

- $NE_{base}$ = The annual natural gas consumption of the base building model (GJ)
- $EE_{retrofitted}$ = The annual electricity consumption of the retrofitted building model (GJ)
- $NE_{retrofitted}$ = The annual natural gas consumption of the retrofitted building model (GJ)
- $EF$ = The local grid electricity emission factor ($tCO_2e/GJ$)
- $NF$ = The local grid natural gas price ($tCO_2e$ /GJ)
- $CT$ = The carbon tax (CAD/$tCO_2e$)

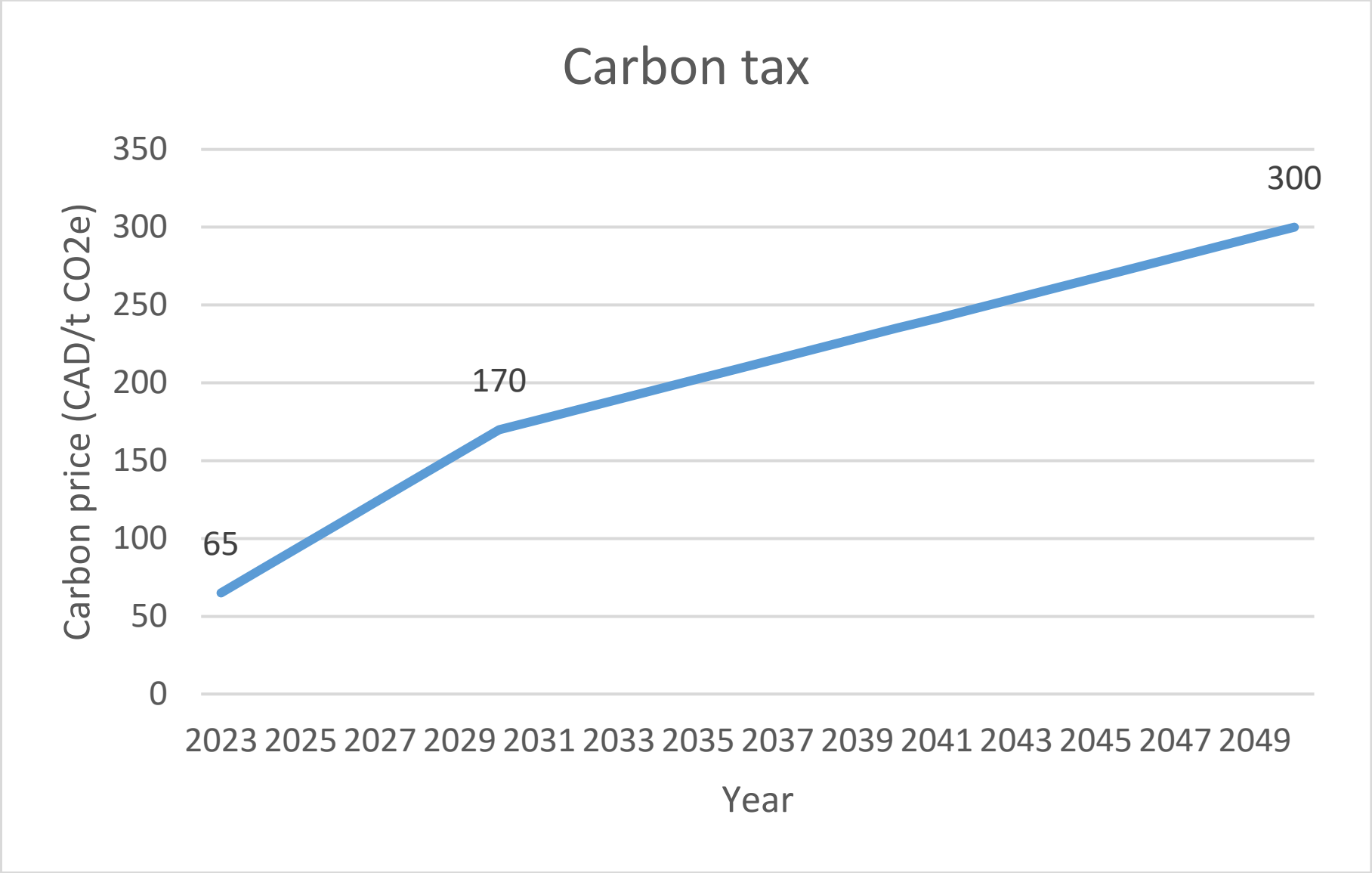

Figure 5. Carbon tax cost

### 4.3 Operational cost

Operational cost of a retrofit has three main components including operational energy cost, maintenance costs, and replacement costs. Reliable maintenance cost data figures associated with different retrofits were not found in the literature. On the other hand, the maintenance costs of residential energy system components are significantly lower compared to operational energy costs associated with the energy system due to energy use. Therefore, only energy cost savings were considered under the operational costs. The energy cost savings and replacement costs were calculated in comparison to the base (existing condition) building using building energy simulations. Energy simulation results can be used to determine the annual operational cost savings of a given retrofit strategy.

$$\Delta AOCS = \left(EE_{base} - EE_{retrofitted}\right) * EP + \left(NE_{base} - NE_{retrofitted}\right) * NP + \Delta CTS$$

Where,

- $\Delta AOCS$ = The annual operational cost savings (CAD)
- $EP$ = The local grid electricity price (CAD/kWh)
- $NP$ = The local grid natural gas price (CAD/GJ)

The net present value (NPV) of the operational cost savings was considered in LCC calculations to account for time value of money. The NPV of the operational cost savings can be calculated using the following equation.

$$\Delta OCS = \sum_{t=0}^{T} \frac{\Delta AOCS}{(1+r)^t}$$

Where,

- $\Delta OCS =$ The net present value of operational cost savings
- $r =$ The discounted rate (%)
- $T =$ The project lifetime

The total life cycle cost of a retrofit measure can be determined by the following equation.

$$LCC = UC - \Delta OCS$$

# 5 Results

## 5.1 Energy simulation results

This section presents the per- and post-retrofit energy performance of the case study building.

### 5.1.1 Performance of the base building models

Table 2. Annual energy, cost and emission impact of the base building models

| Parameter | E-building | NG-building |
|---|---|---|
| Total energy consumption (GJ) | 2125.11 | 2212.47 |
| Energy cost (CAD) | 56.08K | 52.20K |
| GHG emission ($tCO_2e$) | 6.79 | 27.64 |

### 5.1.2 Performance of the selected energy retrofit measures

The section presents the energy performance of the selected energy retrofit measures. Figure 6 depicts the annual energy reductions of individual retrofit measures for the two buildings. NG-buildings present much more cost saving potential compared to E-buildings. Solar PV presents the greatest energy saving potential (around 410 GJ), followed by ASHP (320 GJ). Windows and water HP have similar energy performance, with the energy reduction of around 170 GJ. In terms of upgrades in envelope insulation, wall insulation enhancement can save more energy use compared to roof insulation. Appliances and lighting enhancement show the least energy saving potential among the selected energy retrofit measures.

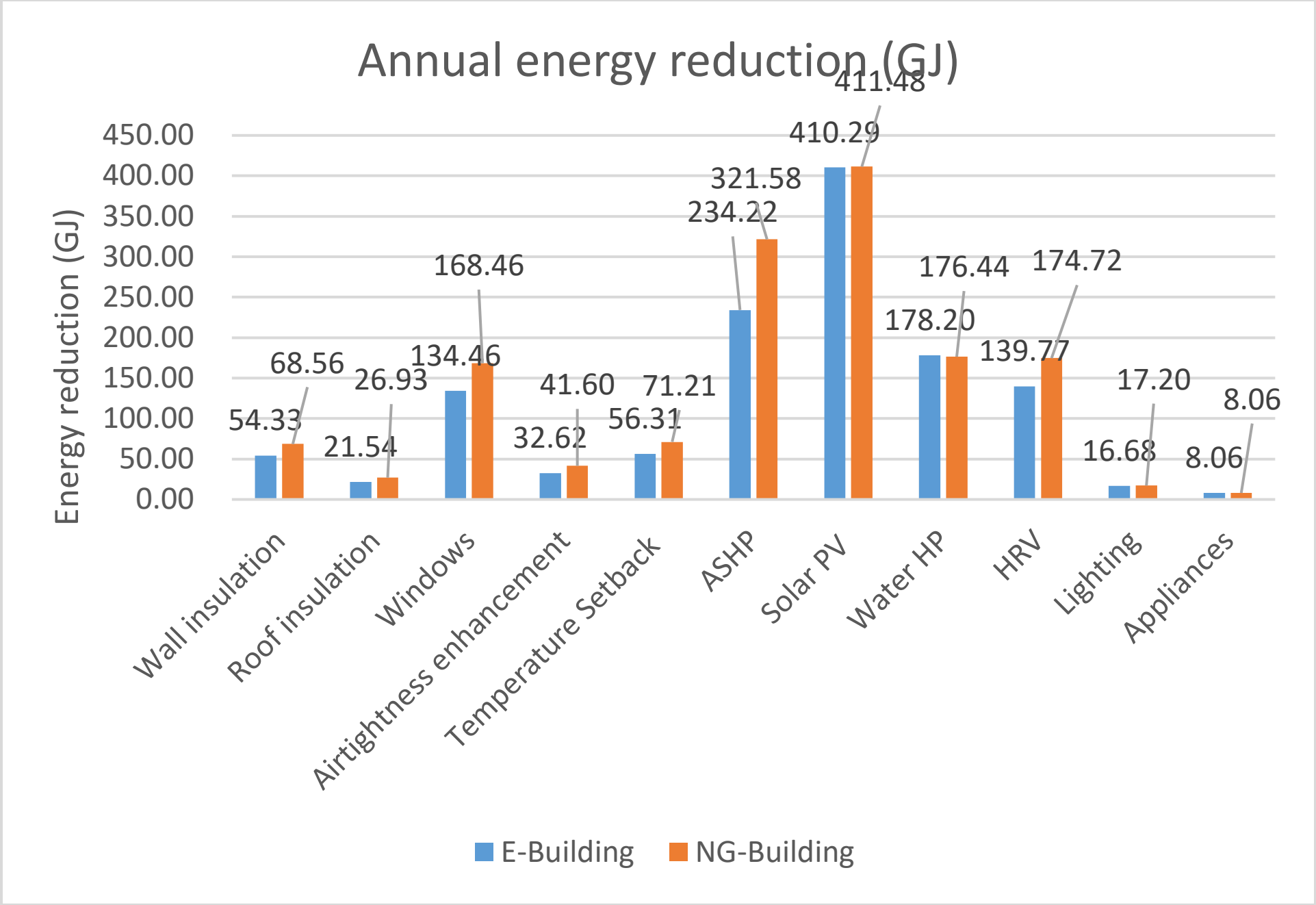

Figure 6. Annual energy reductions of individual retrofit measures

As shown in Figure 7, E-buildings present much more cost saving potential compared to NG-buildings. Similar to energy saving potential, the installation of solar PV is the most cost-effective energy retrofit measure, showing the energy cost saving of around 10K CAD. In addition, ASHP presents significant annual energy cost saving potential for NG-buildings (6.18K CAD). However, the energy cost saving for NG-building is not significant. This because the space heating energy source switches from natural gas to electricity, and the electricity price is higher than natural gas price.

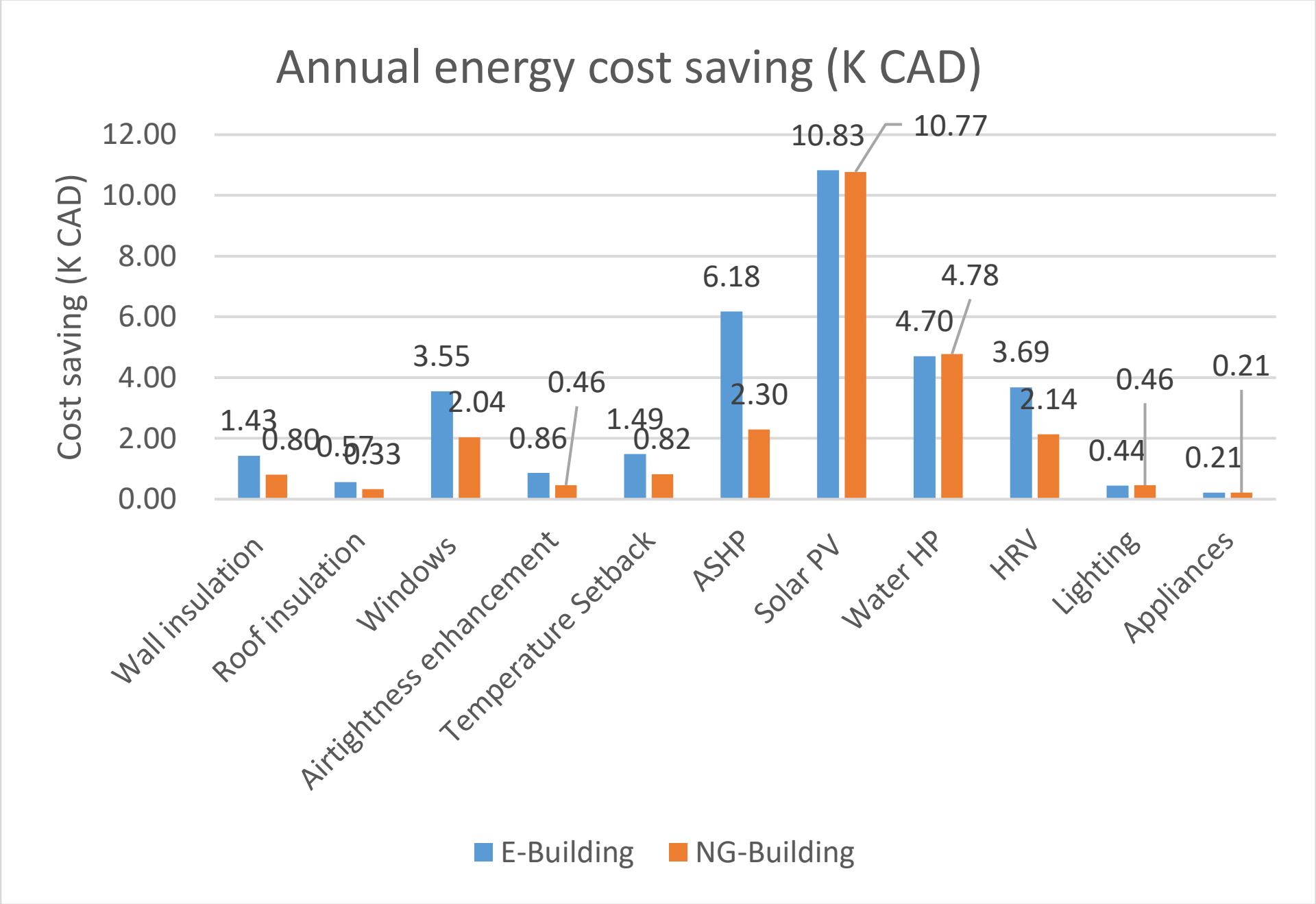

Figure 7. Annual energy cost savings of individual retrofit measures

Figure 8 shows the annual emission reductions of the selected energy retrofit measures. NG-buildings have greater emission reduction potential compared to E-buildings due to a higher emission factor of natural gas. The installation of ASHP in NG-buildings can significantly reduce GHG emissions (21.60 $tCO_2e$), followed by HRV (8.79 $tCO_2e$), and windows (8.55 $tCO_2e$). Upgrades in wall insulation and temperature setback can approximately reduce GHG emissions by 5 t $CO_2e$ for NG-buildings. Other energy retrofit measures are not significant for emission reductions.

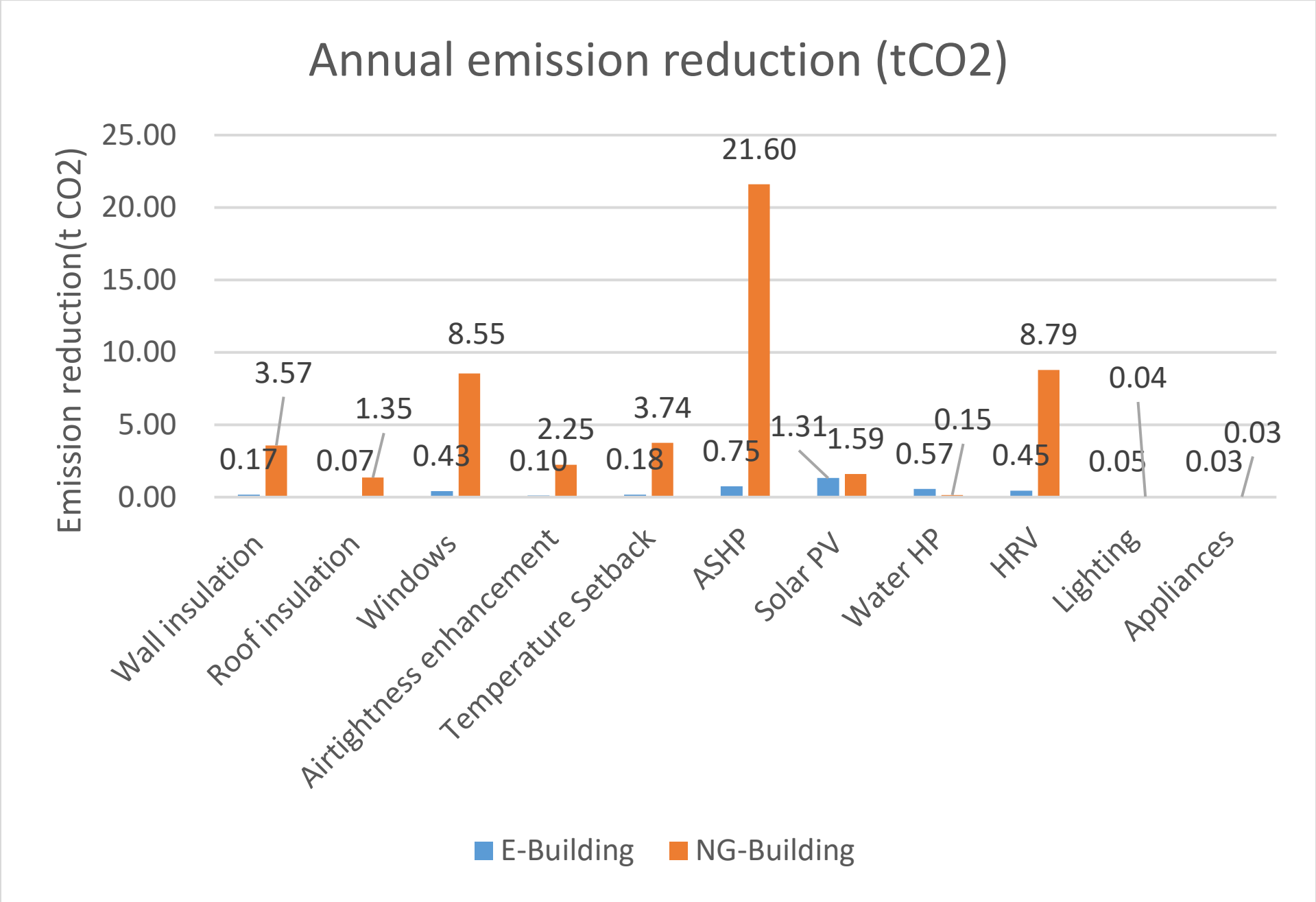

Figure 8. Annual GHG emissions of individual retrofit measures

The following figures present the cumulative effect of annual energy consumptions, energy cost, and GHG emissions for the studied buildings. The energy consumptions can be decreased from around 2200 GJ to 1200 GJ, reduced by 44%. In terms of energy costs, the cost saving potential of E-buildings is greater than that of NG-buildings, with the former decreasing by 35K CAD and the latter decreasing by 27K CAD. However, NG-buildings can produce much more GHG emission reductions compared to E-buildings. The annual GHG emissions of NG-buildings can be decreased from 27.64 tCO2e to 3.77 tCO2e by implementing these retrofit measures. Thus, governments and stakeholders should pay more attention to NG-building to achieve the target of emission reductions.

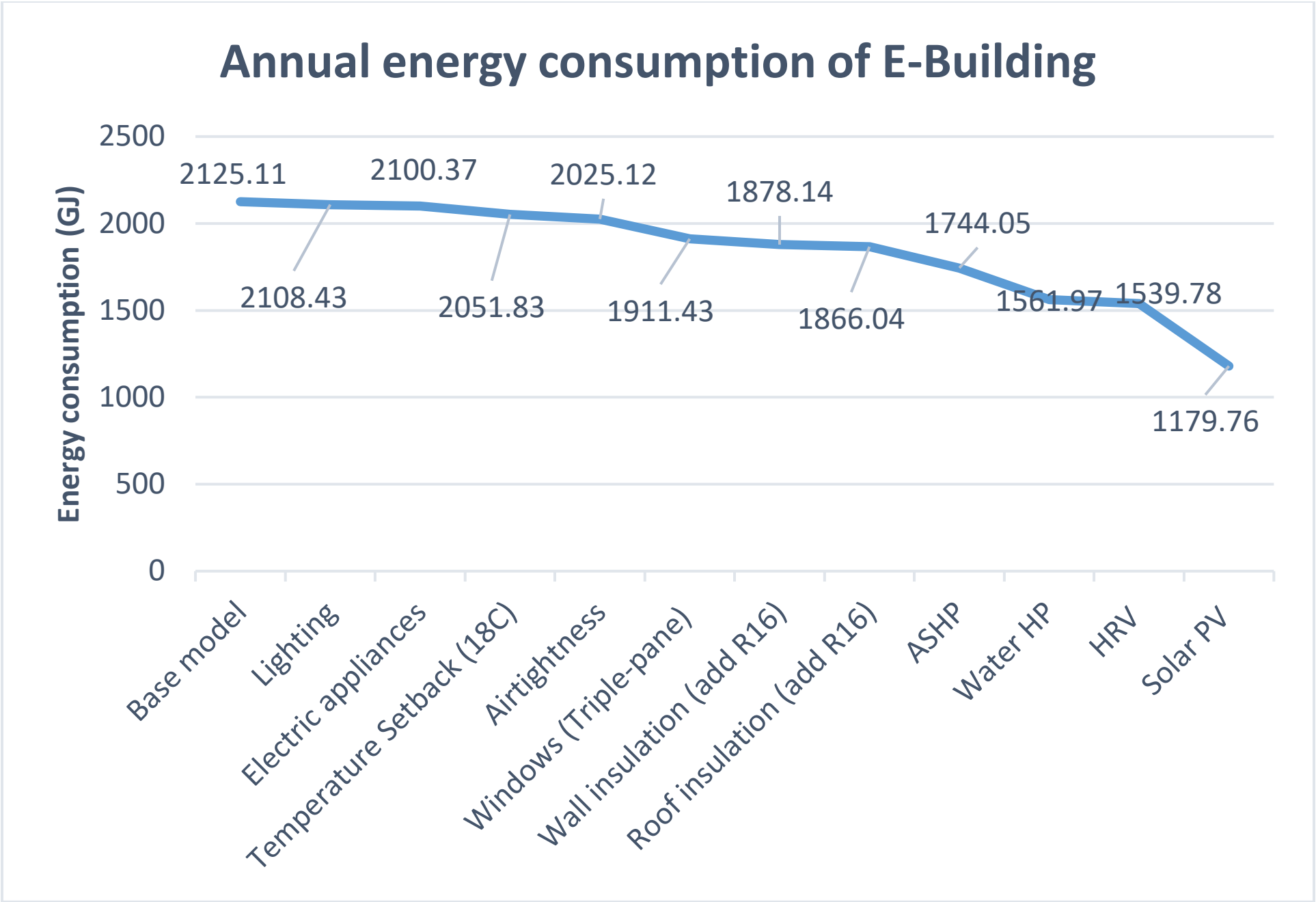

Figure 9. Cumulative effect of energy consumption for E-Building

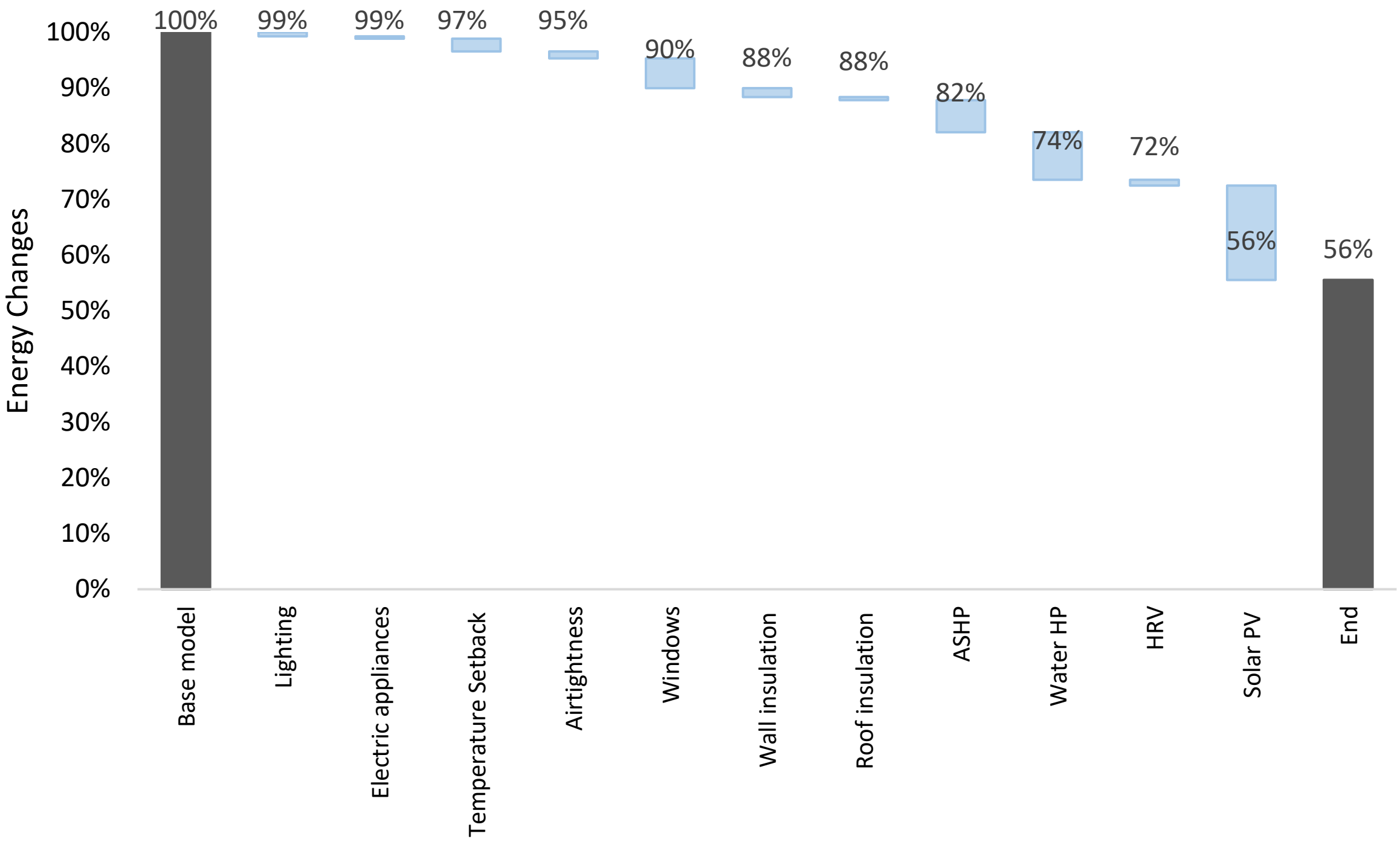

Figure 10. Cumulative effect of energy consumption for E-Building (%)

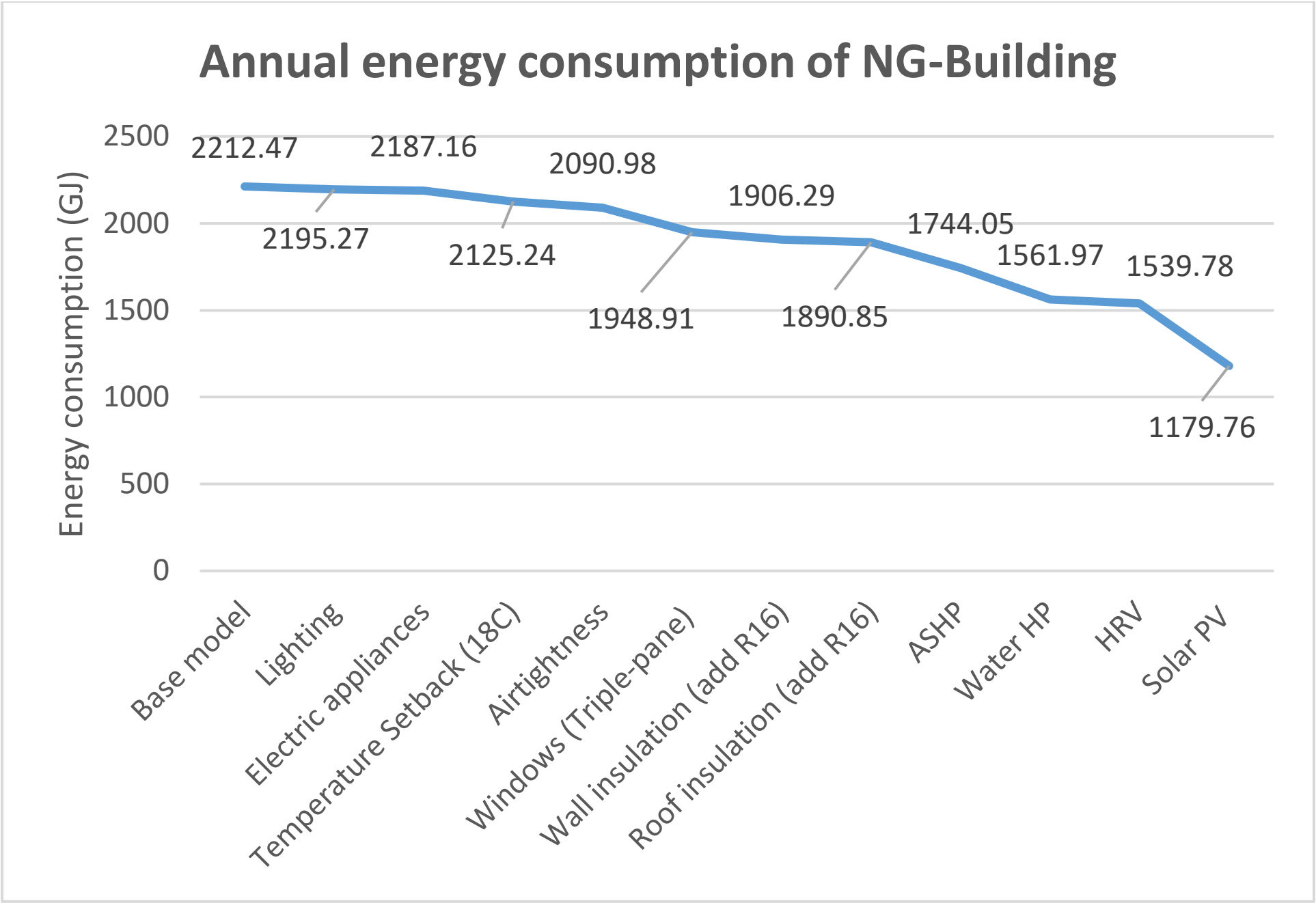

Figure 11. Cumulative effect of energy consumption for NG-Building

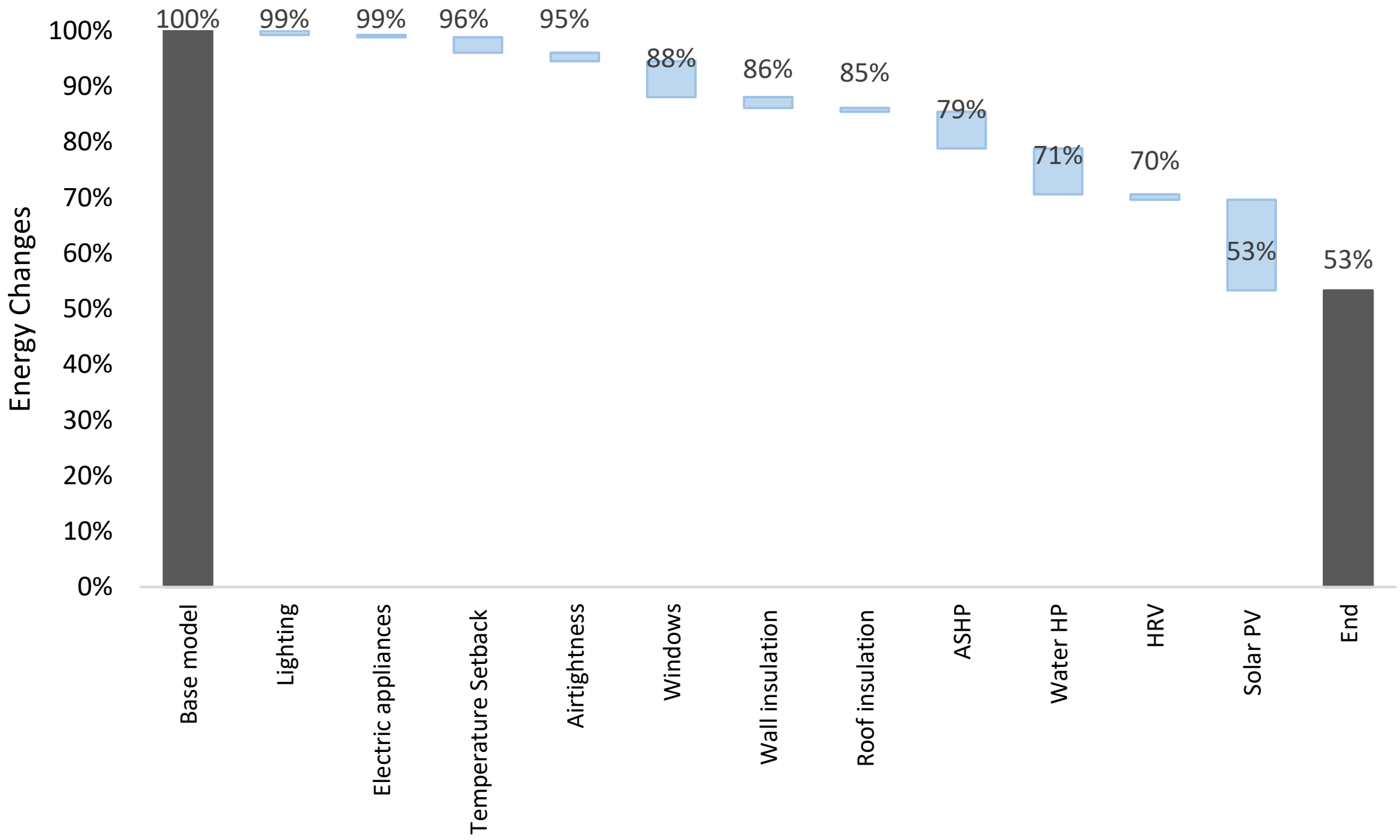

Figure 12. Cumulative effect of energy consumption for NG-Building (%)

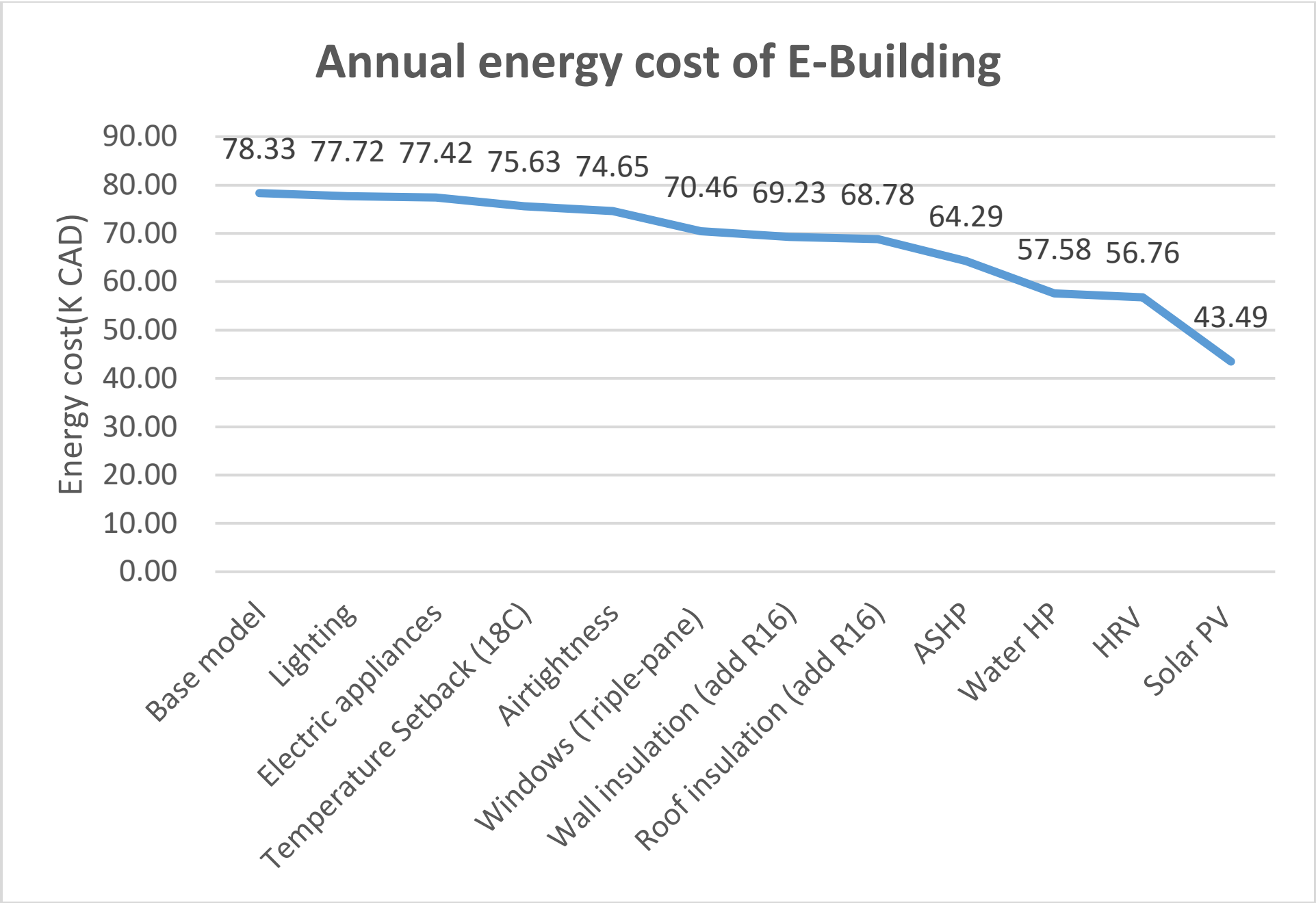

Figure 13. Cumulative effect of energy cost for E-Building

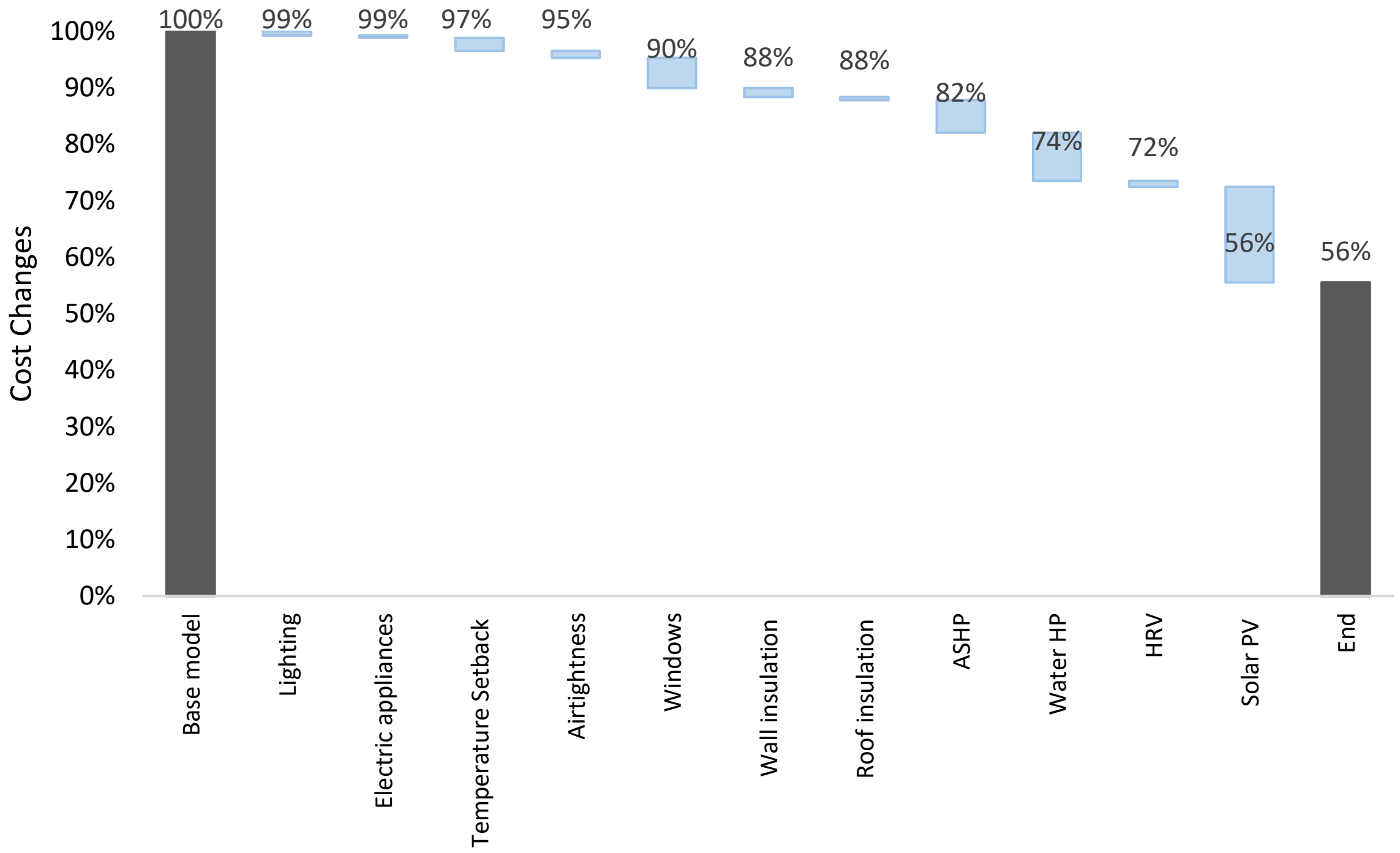

Figure 13. Cumulative effect of energy cost for E-Building (%)

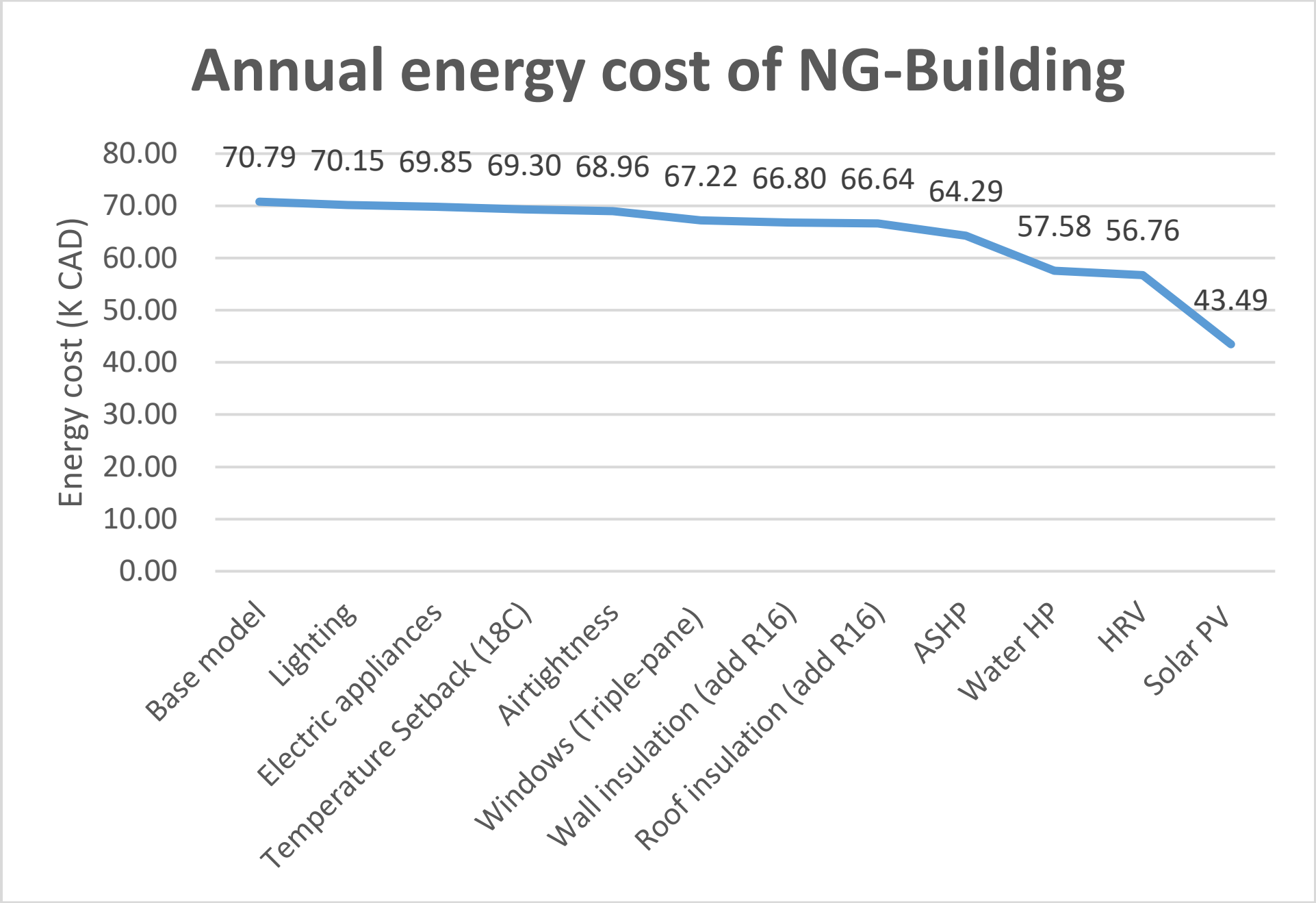

Figure 14. Cumulative effect of energy cost for NG-Building

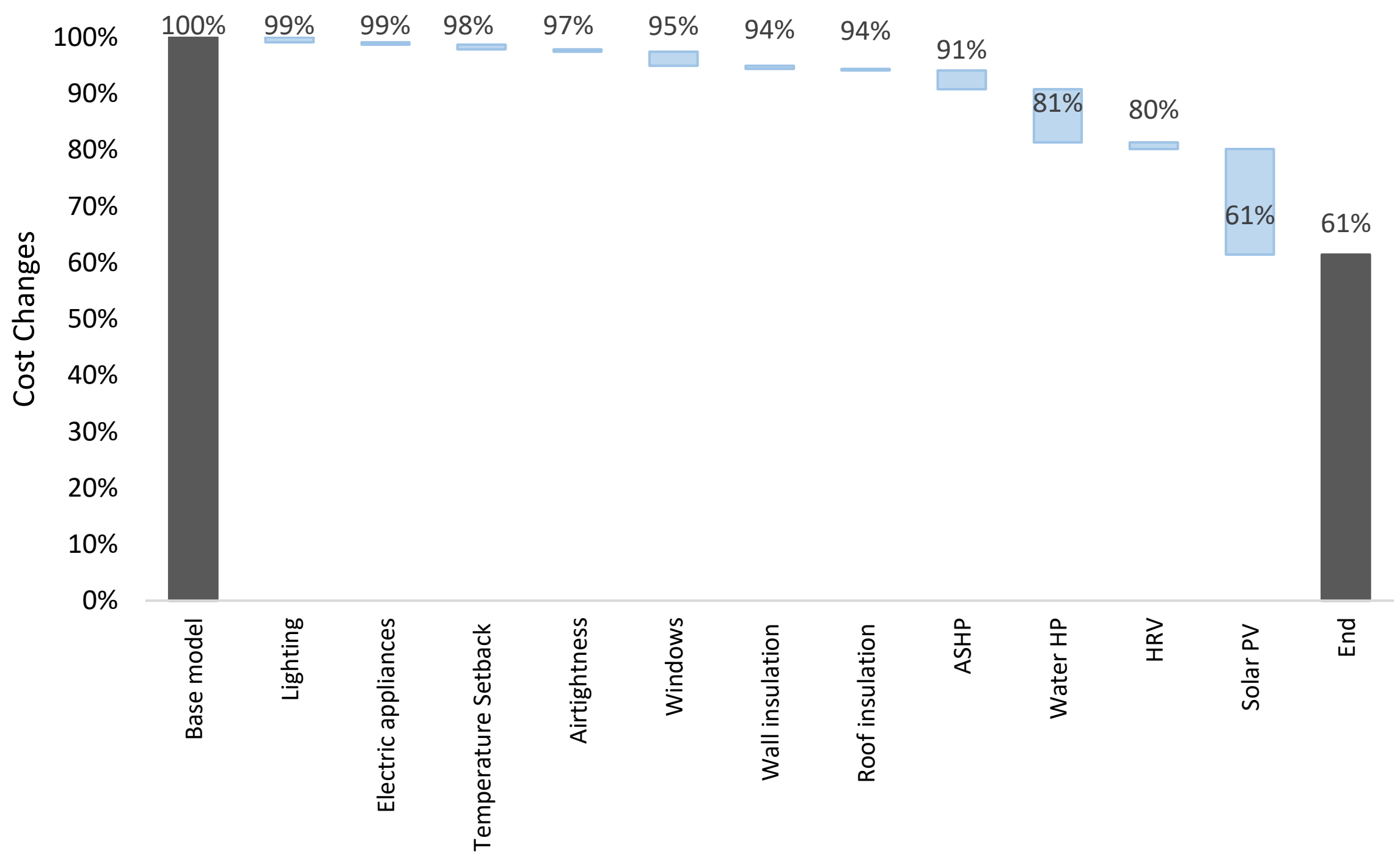

Figure 14. Cumulative effect of energy cost for NG-Building (%)

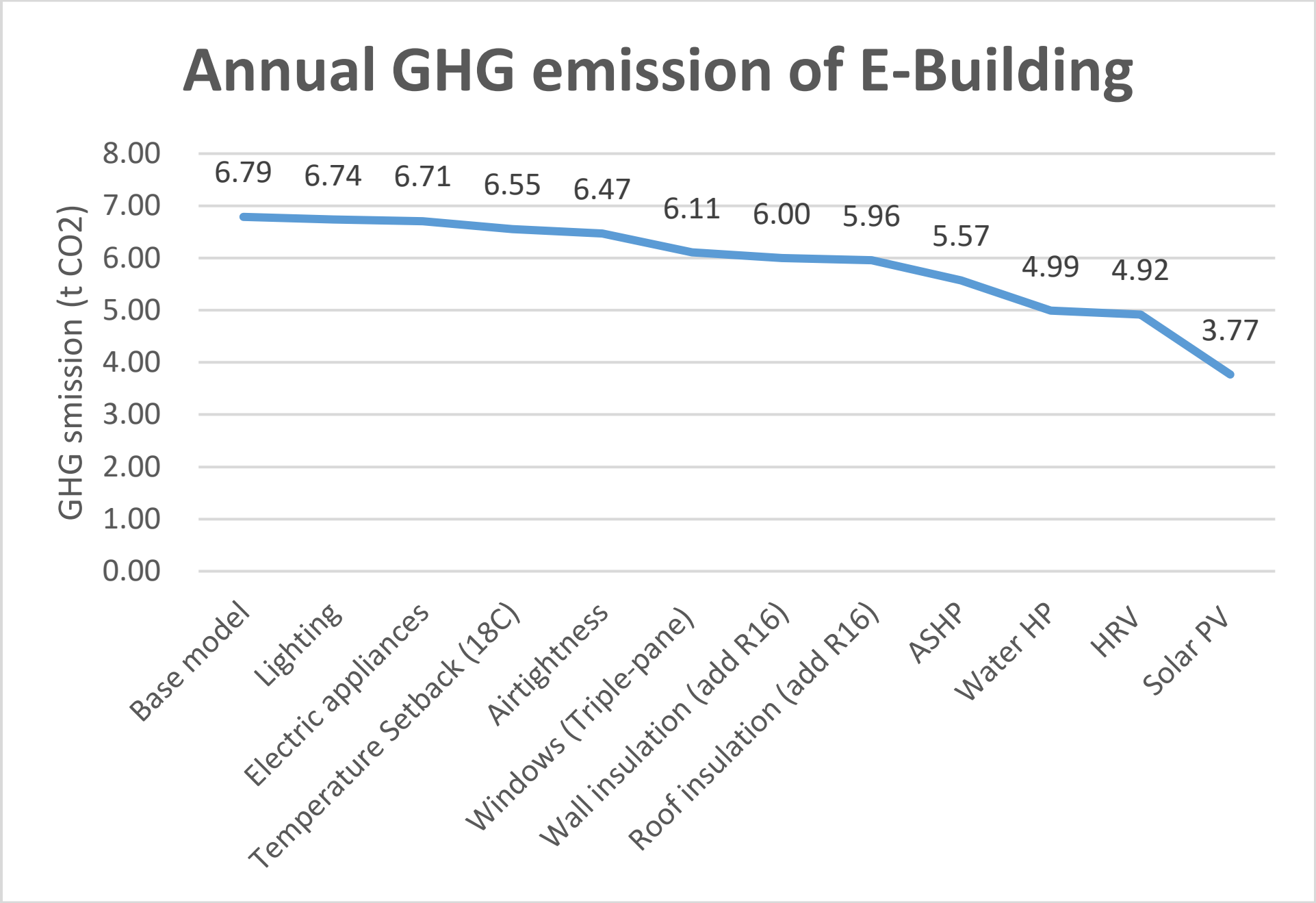

Figure 15. Cumulative effect of GHG emission for E-Building

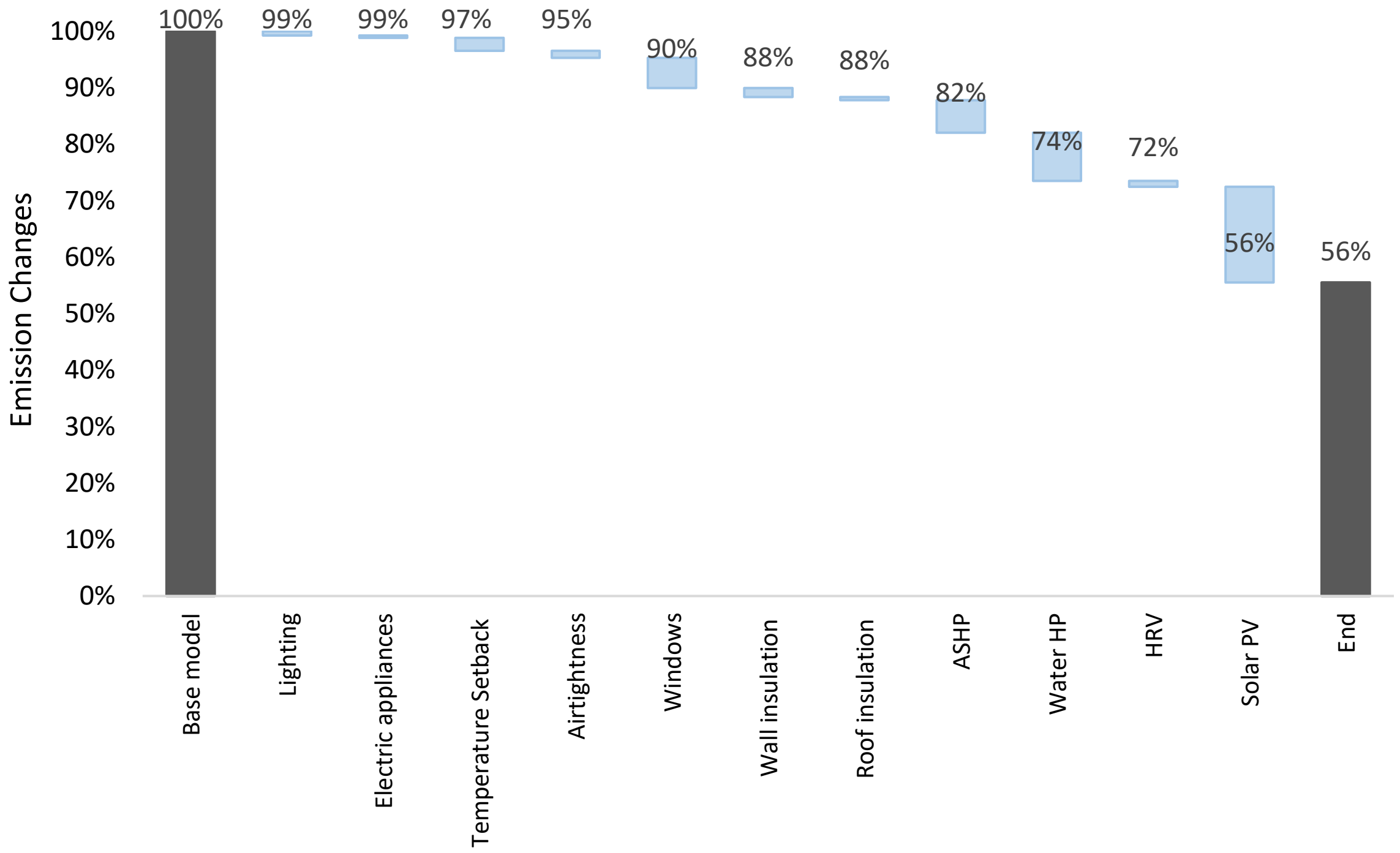

Figure 15. Cumulative effect of GHG emission for E-Building (%)

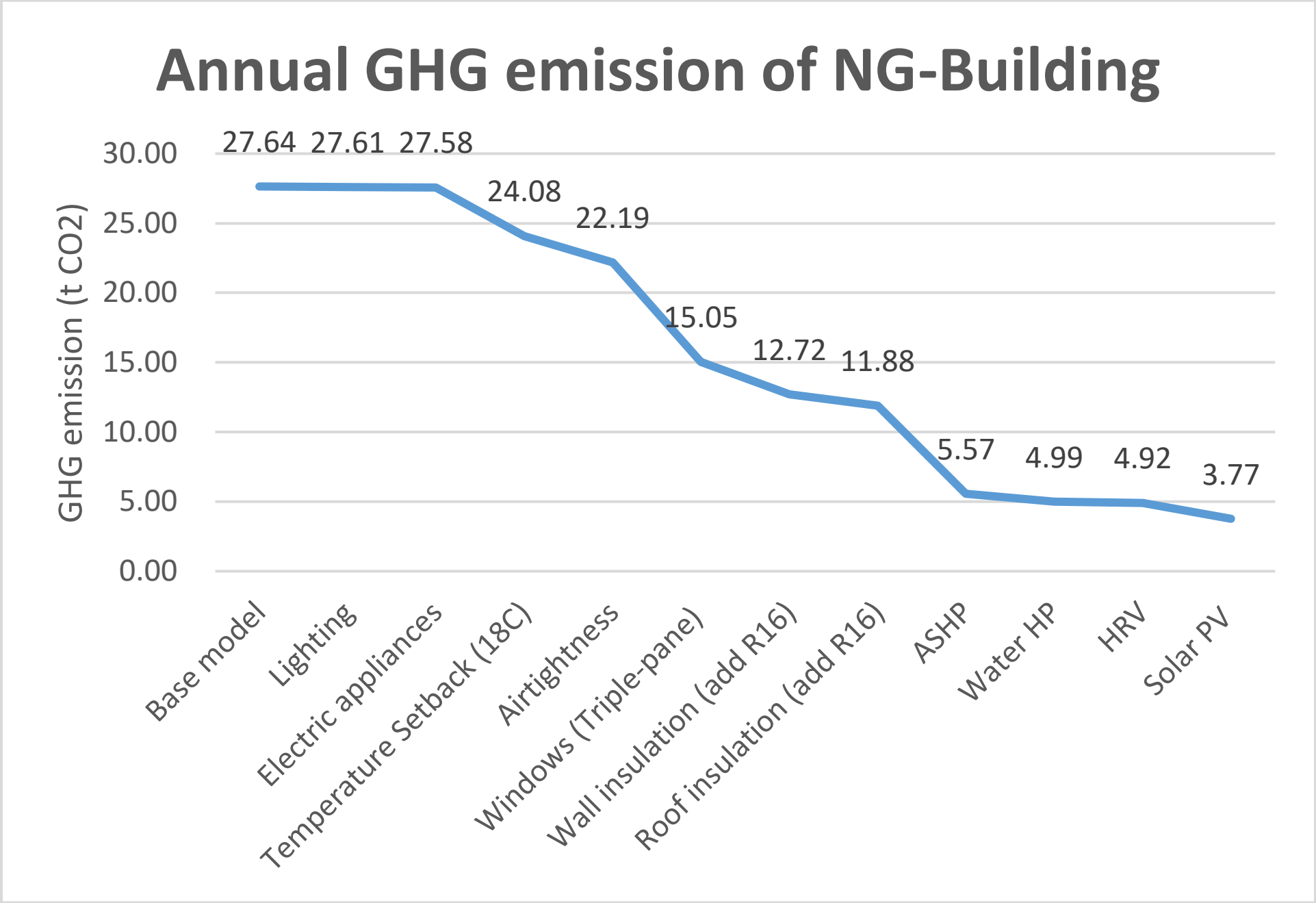

Figure 16. Cumulative effect of GHG emission for NG-Building

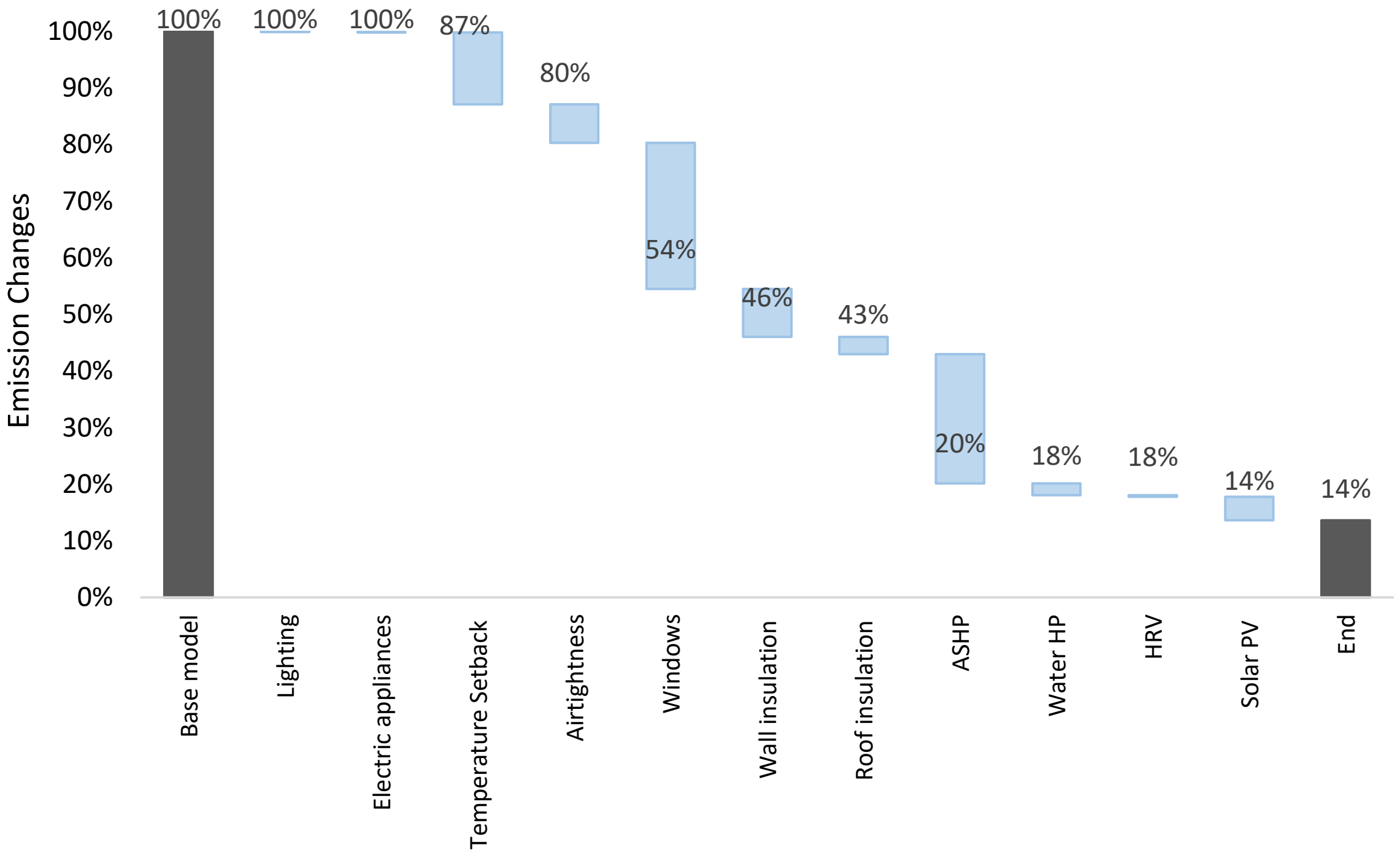

Figure 17. Cumulative effect of GHG emission for NG-Building (%)

## 5.2 Life cycle costing results

This section discusses the life cycle costing results for each of the studied energy retrofit measures. For E-buildings, Figure 18 indicates that the life cycle costs of upgrades in temperature setback is negative, which means that the operational energy cost savings of these retrofit measures are higher than the upfront costs of these measures. Among these retrofit measures, temperature setback presents the greatest cost saving potential, followed by lighting, and airtightness enhancement. On the other hand, upgrades in windows, ASHP, and solar PV are not economical.

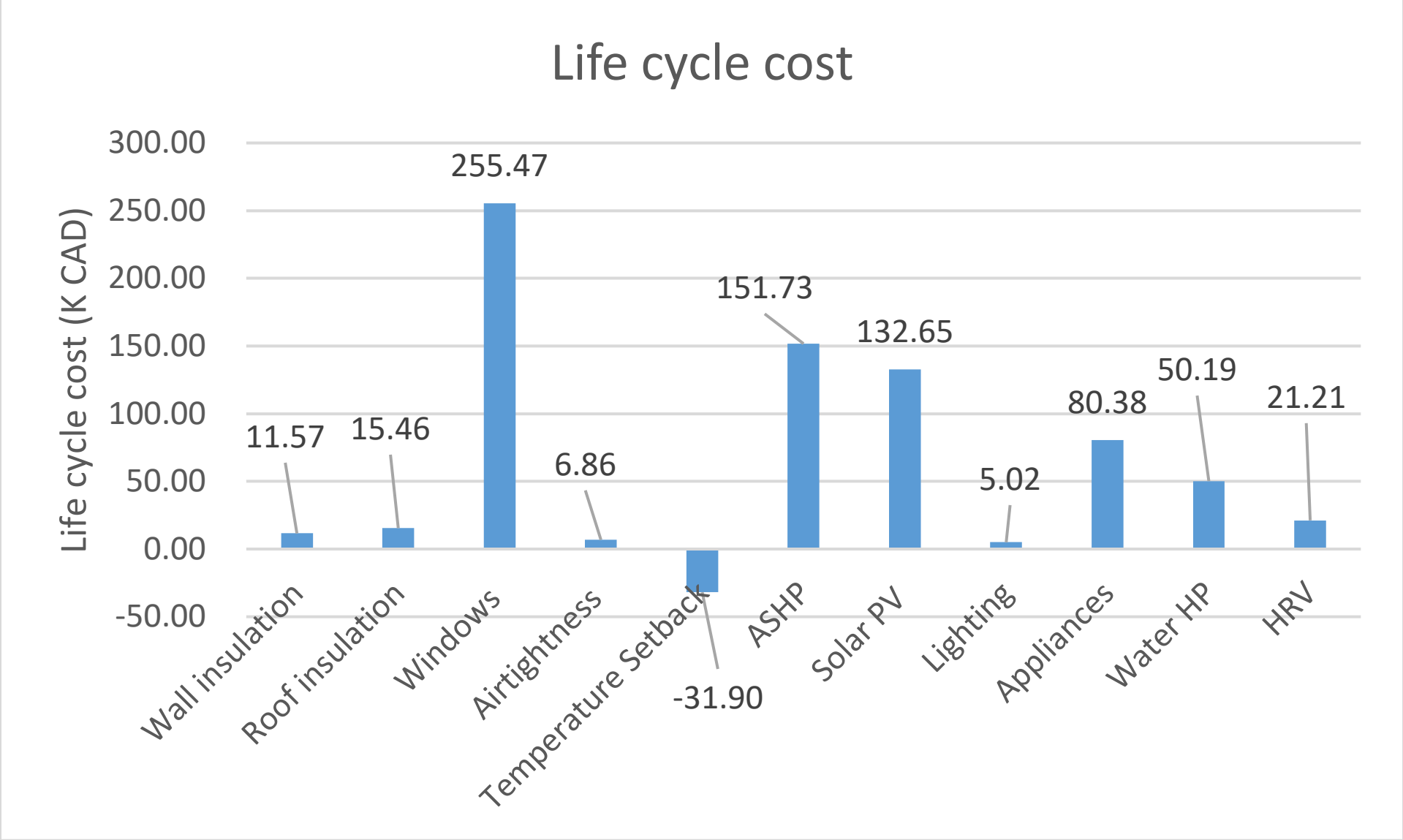

Figure 18. Life cycle cost of individual retrofit measures for E-Building

The life cycle costs of the retrofit measures for NG-buildings are shown in Figure 19. Temperature setback and lighting enhancement are the most cost-effective energy retrofits measures. In addition, airtightness, wall insulation, roof insulation are also economical retrofit measures. However, the replacement of windows can significantly increase life cycle costs (around 270 KCAD). Furthermore, ASHP and solar PV are not economical energy retrofit measures, which can increase life cycle costs of 220K CAD and 120K CAD, respectively.

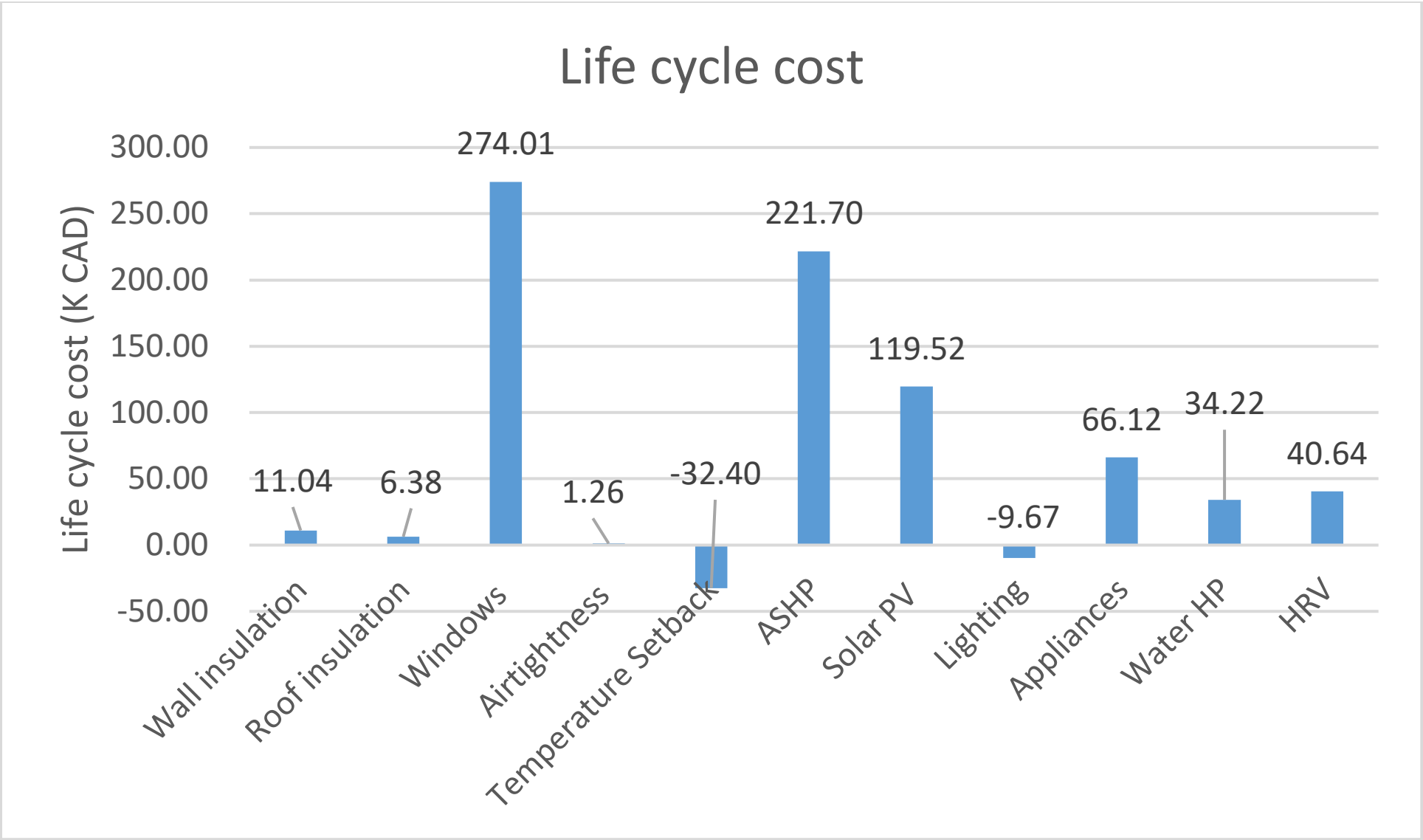

Figure 19. Life cycle cost of individual retrofit measures for NG-Building

## 6 Conclusion

Energy retrofits play an essential role in reduce building energy consumptions and associated GHG emissions. This study employed EnergyPlus to examine the energy performance of 11 energy retrofit measures for typical multi-unit residential buildings located in the City of Richmond, British Columbia, Canada. The energy simulation results were combined with cost and emission impact data to calculate economic and environmental performance of the buildings. The results indicated that NG-buildings can produce significant GHG emission reductions (from 27.64 $tCO_2e$ to 3.77 $tCO_2e$) by implementing these energy retrofit measures. In terms of energy savings, solar PV, ASHP, water heater HP, and HRV enhancement have great energy saving potential compared to other energy retrofit measures. In addition, temperature setback, lighting, and airtightness enhancement present the best economic performance from a life cycle perspective. However, windows, ASHP, and solar PV, are not economical choices because of higher life cycle costs. While ASHP can increase life cycle costs for NG-buildings, with the financial incentives provided by the governments, ASHP could be the best choice to reduce GHG emissions when stakeholders make decisions on implementing energy retrofits.

## References

[1] L. Shen, B. He, L. Jiao, X. Song, and X. Zhang, "Research on the development of main policy instruments for improving building energy-efficiency," *J. Clean. Prod.*, vol. 112, no. 2016, pp. 1789–1803, 2016.

[2] Government of Canada, "Government of Canada sets ambitious GHG reduction targets for federal operations," 2017. .

[3] "Canada's official greenhouse gas inventory - Canada.ca." .

[4] "Energy Step Code – Government of British Columbia." .

[5] "Energy and Greenhouse Gas Emissions (GHGs) | Natural Resources Canada." .

[6] A. Charles, W. Maref, and C. M. Ouellet-Plamondon, "Case study of the upgrade of an existing office building for low energy consumption and low carbon emissions," *Energy Build.*, vol. 183, pp. 151–160, Jan. 2019.

[7] B. Palacios-Munoz, B. Peuportier, L. Gracia-Villa, and B. López-Mesa, "Sustainability assessment of refurbishment vs. new constructions by means of LCA and durability-based estimations of buildings lifespans: A new approach," *Build. Environ.*, vol. 160, p. 106203, Aug. 2019.

[8] Z. J. Ma, P. Cooper, D. Daly, and L. Ledo, "Existing building retrofits: Methodology and state-of-the-art," *Energy Build.*, vol. 55, pp. 889–902, 2012.

[9] R. Ruparathna, K. Hewage, and R. Sadiq, "Economic evaluation of building energy retrofits: A fuzzy based approach," *Energy Build.*, vol. 139, pp. 395–406, 2017.

[10] H. Feng, D. R. Liyanage, H. Karunathilake, R. Sadiq, and K. Hewage, "BIM-based life cycle environmental performance assessment of single-family houses: Renovation and reconstruction strategies for aging building stock in British Columbia," *J. Clean. Prod.*, vol. 250, p. 119543, Dec. 2020.

[11] J. Weiss, E. Dunkelberg, and T. Vogelpohl, "Improving policy instruments to better tap into homeowner refurbishment potential: Lessons learned from a case study in Germany," *Energy Policy*, vol. 44, pp. 406–415, May 2012.

[12] Z. Ma, P. Cooper, D. Daly, and L. Ledo, "Existing building retrofits: Methodology and state-of-the-art," *Energy Build.*, vol. 55, pp. 889–902, 2012.

[13] H. Karunathilake, T. Prabatha, K. Hewage, and R. Sadiq, "Costs of green residences in Canada: An economic and environmental analysis of developing renewable powered building clusters," in *Proceedings, Annual Conference - Canadian Society for Civil Engineering*, 2019.

[14] P. Penna, A. Prada, F. Cappelletti, and A. Gasparella, "Multi-objectives optimization of Energy Efficiency Measures in existing buildings," *Energy Build.*, vol. 95, pp. 57–69, 2015.

[15] Y. Shao, P. Geyer, and W. Lang, "Integrating requirement analysis and multi-objective optimization for office building energy retrofit strategies," *Energy Build.*, vol. 82, pp. 356–368, 2014.

[16] H. Karunathilake, K. Hewage, and R. Sadiq, "Opportunities and challenges in energy

demand reduction for Canadian residential sector: A review," *Renew. Sustain. Energy Rev.*, 2017.

[17] S. Finnegan, C. Jones, and S. Sharples, "The embodied CO2e of sustainable energy technologies used in buildings: A review article," *Energy Build.*, vol. 181, pp. 50–61, 2018.

[18] H. Karunathilake, K. Hewage, and R. Sadiq, "Opportunities and challenges in energy demand reduction for Canadian residential sector : A review," *Renew. Sustain. Energy Rev.*, vol. 82, no. February 2017, pp. 2005–2016, 2018.

[19] T. Prabatha, K. Hewage, H. Karunathilake, and R. Sadiq, "To retrofit or not? Making energy retrofit decisions through life cycle thinking for Canadian residences," *Energy Build.*, vol. 226, 2020.

[20] H. Zhang, K. Hewage, T. Prabatha, and R. Sadiq, "Life cycle thinking-based energy retrofits evaluation framework for Canadian residences: A Pareto optimization approach," *Build. Environ.*, vol. 204, no. April, p. 108115, 2021.

[21] H. Zhang, H. Feng, K. Hewage, and M. Arashpour, "Artificial Neural Network for Predicting Building Energy Performance: A Surrogate Energy Retrofits Decision Support Framework," *Buildings*, vol. 12, no. 6, p. 829, 2022.

[22] S. A. Sharif and A. Hammad, "Simulation-Based Multi-Objective Optimization of institutional building renovation considering energy consumption, Life-Cycle Cost and Life-Cycle Assessment," *J. Build. Eng.*, vol. 21, no. June 2018, pp. 429–445, 2019.

[23] M. Heidari, M. H. Rahdar, A. Dutta, and F. Nasiri, "An energy retrofit roadmap to net-zero energy and carbon footprint for single-family houses in Canada," *J. Build. Eng.*, vol. 60, no. May, p. 105141, 2022.

[24] H. Zhang, K. Hewage, H. Karunathilake, H. Feng, and R. Sadiq, "Research on policy strategies for implementing energy retrofits in the residential buildings," *J. Build. Eng.*, vol. 43, no. August, p. 103161, 2021.

[25] H. Feng, J. Zhao, H. Zhang, S. Zhu, and D. Li, "Uncertainties in whole-building life cycle assessment : A systematic review," *J. Build. Eng.*, vol. 50, no. February, p. 104191, 2022.